\documentclass[prd,twocolumn,superscriptaddress,showpacs]{revtex4}

\usepackage{graphicx}
\usepackage{latexsym}
\usepackage{bm}
\usepackage{amsmath}

\newcommand{\comment}[1]{}

\newcommand{\be}{\begin{eqnarray}}
\newcommand{\ee}{\end{eqnarray}}
\newcommand{\munu}{\mu\nu}
\newcommand{\FF}{{\overline F}}
\newcommand{\rr}{{\overline \rho}}

\begin{document}


\title{Classical Gluodynamics of High Energy Nuclear Collisions:
 an Erratum and an Update}

\author{Alex Krasnitz}
\affiliation{FCT and CENTRA, Universidade do Algarve, Campus de Gambelas,
   P-8000 Faro, Portugal}
\author{Yasushi Nara}
\affiliation{%
 Physics Department, University of Arizona, Tucson, AZ, 85721.
}
\author{Raju Venugopalan}
\affiliation{%
\ Physics Department, Brookhaven National Laboratory, Upton, N.Y. 11973,
U.S.A.
}
\affiliation{%
 RIKEN BNL Research Center, Brookhaven National Laboratory,
                Upton, N.Y. 11973, U.S.A.
}

\date{\today}

\begin{abstract}

We comment on the relation of our previous work on the classical
gluodynamics of high energy nuclear collisions to recent work by Lappi
(hep-ph/0303076). While our results for the non-perturbative number
liberation coefficient agree, those for the energy disagree by a
factor of 2. This discrepancy can be traced to an overall
normalization error in our non-perturbative formula for the energy. 
When corrected for, all previous
results are in excellent agreement with those of Lappi. The
implications of the results of these two independent computations for
RHIC phenomenology are noted.
 
\end{abstract}


\maketitle

\section{Introduction}

In a series of recent papers, we computed numerically the classical
gluodynamics of the early stages of very high energy nuclear
collisions. These computations were first performed for an SU(2)
Yang-Mills gauge theory~\cite{AR99,AR00,AR01} and later for the SU(3)
Yang-Mills gauge theory~\cite{AYR01,AYR02a,AYR02b}.  In a very recent
paper, Lappi has independently performed the same computation for the
SU(3) gauge theory~\cite{Lappi}. We will show that the apparent
discrepancies in the two computations can be understood as arising
from an incorrect overall normalization in the non-perturbative formula 
for the energy in our previous work. Since the
problem arises from a normalization error, all of our previous results
are still useful if interpreted appropriately. The purpose of this
note is primarily to clarify the sources of discrepancy in the two
computations and to present corrected results wherever necessary. Now
that two independent computations can be shown to agree, we will show
that they significantly constrain phenomenological interpretations of
the RHIC data.

This note is organized as follows. In Section II, we discuss the
different conventions for the small x classical effective Lagrangean density used
in the literature and isolate the source of the normalization error in
our previous computations. A reader uninterested in the details can skip 
right ahead to Section III where we discuss the corrected results for
various quantities.  The implications of these results
for RHIC phenomenology are discussed in Section IV.

\section{Normalization of Classical Effective Lagrangean}

There are two conventions for the classical effective Lagrangean density~\cite{MV} that are followed in the 
literature. This is occasionally confusing so we shall discuss these explicitly here.
The first of these is 
\be
L_{I} [A,\rho] = {1\over 4}\, F^a F^a  - \rho^a A^a + i {\rho^a \rho^a 
\over 2\mu_A^2} \, .
\label{S1}
\ee
For convenience, all Lorentz indices and space--time integrals have
been suppressed here. For example, $F^a$ denotes the field strength tensor. 
In this form of the
effective Lagrangean density, the coupling constant $g$ is absorbed in the
definition of the charge density $\rho^a$.
The color charge squared per unit
area $\mu_A^2$ contains a factor $g^2$. The correlator of color
charges is then
\be
\langle\rho^a (x_\perp)\rho^a (y_\perp)\rangle = \mu_A^2\, \delta^{ab}\, 
\delta^{(2)}(x_\perp - y_\perp) \, .
\label{C1}
\ee
This form of the effective Lagrangean density is used in several works -- for 
instance, Refs.~\cite{KovMuell,IancuMcLerran}.

The other form of the effective Lagrangean density found in the
literature, for instance in Refs.~\cite{MV,JKW,GyulassyMcLerran}
is 
\be
L_{II}[A,\rho] = {1\over 4}\, F^a F^a  - g{\tilde \rho}^a A^a + 
i {{\tilde \rho}^a {\tilde \rho}^a \over 2\mu_N^2} \, .
\label{S2}
\ee
Here $g$ is not absorbed in the definition of the charge density
$\rho^a$ but appears explicitly in the second term of the classical
effective Lagrangean density. The relation between the color charge
squared values in the two Lagrangean densities is therefore 
\be
\mu_A^2 = g^2 \mu_N^2 \, . \nonumber
\label{C2}
\ee
The form of the effective Lagrangean density in Eq.~(\ref{S2}) is the
form used in our papers~\cite{AR99,AR00,AR01,AYR01,AYR02a,AYR02b}, and
the scale we use, $\Lambda_s^2$ is defined to be $\Lambda_s^2 = g^4 \mu_N^2$.
Eq.~(1) of Ref.~\cite{Lappi} has the form
\be 
\langle\rho^a (x_\perp)\rho^a (y_\perp)\rangle = g^2 \mu_N^2\, \delta^{ab}\, 
\delta^{(2)}(x_\perp - y_\perp) \, .
\ee
Thus from Eqs.~(\ref{C1}) and ~(\ref{C2}), it is clear that 
the form of the effective Lagrangean density used by Lappi is 
Eq.~(\ref{S1}), albeit with the color charge squared defined to 
be identical to ours.

In Eq.~(\ref{S1}) and (\ref{S2}), the field strength tensor has the canonical form
\be
F_{\munu}^a = \partial_\mu A_\nu^a -\partial_\nu A_\mu^a + g f^{abc} A_\mu^b A_\nu^c \, .
\ee
In our numerical work -- see for instance Ref.~\cite{AR99} -- the field strength tensor is 
defined as 
\be
{\FF}_{\munu}^a=\partial_\mu {\overline A}_\nu^a -\partial_\nu {\overline A}_\mu^a + 2 f^{abc} 
{\overline A}_\mu^b {\overline A}_\nu^c \, .
\ee
Performing the field re-definition ${\overline A}_\mu^a = {g\over 2} A_\mu^a$, we obtain
\be
{\FF}_{\munu}^a = {g\over 2}\, F_{\munu}^a \, .
\label{FF1}
\ee
The ``canonical'' effective Lagrangean density in Eq.~(\ref{S2}) can then be re-written as 
\be
L_{II} &=& \left( {2\over g}\right)^2\, \left( {1\over 4}\, {\FF}^a {\FF}^a + 
{g^2 \over 2}\,{\tilde \rho}^a 
{\overline A}^a\right)
\nonumber \\
&+& i{{\tilde \rho}^a {\tilde \rho}^a \over 2\mu_N^2} \, .
\ee
With a further re-definition ${\rr}^a = g^2 {\tilde \rho}^a/2$, 
we obtain
\be
L_{(II)}&=& {4\over g^2}\, \left( {1\over 4}\, \FF^a \FF^a + {\rr}^a 
{\overline A}^a\right)
\nonumber \\
&+& i{{\rr}^a {\rr}^a \over 2\,g^4 {\overline \mu}_N^2} \, ,
\label{FF2}
\ee
where ${\overline \mu}_N = \mu_N/2$. With the field re-definition, 
$A^a\rightarrow A^a/g$, both $L_{I}$ and $L_{II}$, as defined in Eqs.~(\ref{S1}) and (\ref{S2}) 
respectively, can be expressed in the  same form as the right hand side of Eq.~(\ref{FF2}),
albeit without the factor of $4$ in front of the first two terms,
and with $\mu_N$ instead of ${\overline \mu}_N$. Since the continuum limit of the lattice Hamiltonian 
in Ref.~\cite{Lappi} corresponds to $L_{I}$ with $A^a\rightarrow A^a/g$ and $\rho^a={\rr}^a/g$, there is 
this overall difference of a factor of 4 in the normalization between our lattice Hamiltonian and that of Lappi 
plus the requirement that the charge densities are related by ${\overline \mu}_N = \mu_N/2$. 

Now consider the expression for the energy per unit rapidity per unit area,
\be
\varepsilon \tau = {1\over 2}\, \left(E^a E^a + B^a B^a\right) \, ,
\label{E1}
\ee
again suppressing Lorentz indices. From Eq.~(\ref{FF1}), 
\be
{\overline B}^a ={g\over 2}\, B^a\,\,\,;\,\,\, {\overline E}^a = {g\over 2}\, E^a \, ,
\ee
which gives 
\be
\varepsilon \tau = {4 \over g^2}\, {\overline \varepsilon} \tau \, .
\label{E2}
\ee
In Refs.~\cite{AR00} and ~\cite{AYR01}, we computed 
\be
{\overline \varepsilon} \tau = {\overline f}_E \, 
\left(g^2 {\overline \mu}_N\right)^3 \, ,
\ee
where ${\overline f}_E$ was determined by an extrapolation of the lattice results to the continuum limit. 
Then from Eq.~(\ref{E1}) and (\ref{E2}), we obtain 
\be
\varepsilon \tau = {4\over g^2}\, {\overline f}_E\, 
\left(g^2{\overline \mu}_N\right)^3 \, .
\ee
However, ${\overline \mu}_N = \mu_N/2$. Performing this substitution, we obtain
\be
\varepsilon \tau = {1\over g^2}\, f_E\, \Lambda_s^3 \, ,
\label{F1}
\ee
where $f_E = {\overline f}_E/2$. Thus the
physical value of the energy liberation coefficient is one half that
computed previously. It accounts precisely for the result found by
Lappi.

The number liberation coefficient remains the same after the field re-definitions. This is because 
the number per unit rapidity per unit area is defined to be 
\be
N = {4\over g^2} {\overline f}_N \left(g^2 {\overline \mu}_N\right)^2 \, .
\ee
Since ${\overline \mu}_N = \mu_N/2$, this replacement here exactly cancels the extra factor of $4$ appearing 
in the normalization. Hence ${\overline f}_N = f_N$ -- our previous result is in agreement with the result of Lappi.

The gluon number distributions in the two papers coincide if we 
replace $\Lambda_s\rightarrow \Lambda_s/2$. The 
resulting comparison of the result of Lappi to our prior result in Ref.~\cite{AYR01} is excellent and is 
shown in Fig.~\ref{fig:dndkt}.

\section{Corrected Results}

Now that we understand the origin of the apparent discrepancies in the numerical results of Ref.~\cite{Lappi} and 
our prior results, we will discuss the corrected results here. As discussed previously~\cite{AR00}, all dimensional 
quantities in the classical effective theory can be expressed in terms of the appropriate power of $\Lambda_s$ times 
a non-perturbative function of $\Lambda_s R$. This is because $\Lambda_s$ and $R$ are the only dimensional scales in 
the problem. $\Lambda_s$ is, in principle, a function of the energy, the centrality and the atomic number. For RHIC 
energies, one can broadly estimate that $\Lambda_s\sim 1$-2 GeV (more on this in the next section). The radius of a 
gold nucleus is approximately $6.5$ fm, so $\Lambda_s R$ lies in the range 
$32.5$ -- $65$ for Au-Au collisions 
at RHIC. We obtain, for the transverse energy,
\be
{1\over \pi R^2}\,{dE_T\over d\eta} = {f_E(\Lambda_s R)\over g^2}\, \Lambda_s^3 \, .
\label{G1}
\ee
Previously, in Ref.~\cite{AYR01}, we had $f_E(25)=0.537$ and
$f_E(83.7)=0.497$. Following Eq.~(\ref{F1}), this should be corrected to
read $f_E(50)=0.27$ and $f_E(167.4)=0.25$. The corresponding
expression for the gluon number is
\be
{1\over \pi R^2}\,{dN\over d\eta} = {f_N(\Lambda_s R)\over g^2}\, \Lambda_s^2 \, ,
\label{G2}
\ee
with $f_N(50)=0.315$ and $f_N(167.4)=0.3$.
 The variation in the number with $\Lambda_s R$ 
is very small, on the order of 10\%. 
These results are in good agreement with those of Lappi. 

\begin{figure}[t]
\includegraphics[width=3.3in]{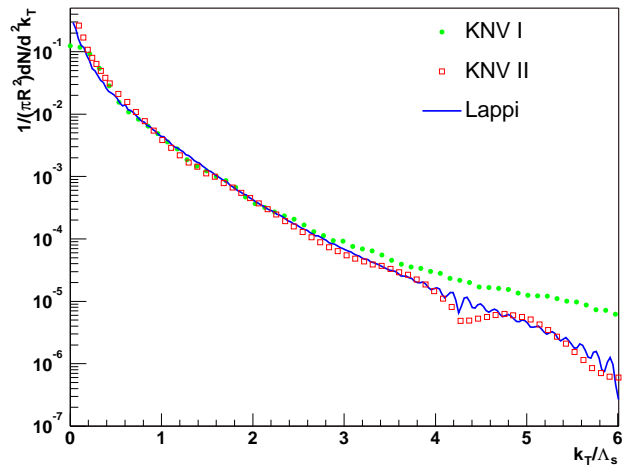}
\caption{Comparison of gluon transverse momentum distributions
per unit area
as a function of $k_{T}/\Lambda_s$.
KNV I (circles): 
the number defined as in
Eq.~(\ref{eq:gn1}),
with $k_T$ taken to mean the lattice wave number
along one of the principal directions.
KNV II (squares) and Lappi (solid line):
the number defined as in
Eq.~(\ref{eq:gn3}),
 with an average over the entire Brillouin zone and with $k_T$ taken to 
mean the frequency $\omega({\vec k}_T)$. 
}
\label{fig:dndkt}
\end{figure}

As we have explained in detail in our earlier papers, there are a variety
of ways of defining the particle number for a generic field $\phi$ and its 
conjugate momentum $\pi$. Let us compare the results for the following three
definitions (here and in the following $k_T$ is the transverse momentum of
the field):
\begin{eqnarray}
 n(\vec k_T) &=& \sqrt{\langle|\phi(\vec k_T)|^2\rangle\langle|\pi(\vec k_T)|^2\rangle},
          \label{eq:gn1}\\
 \ &=& \omega(\vec k_T) \langle|\phi(\vec k_T)|^2\rangle \label{eq:gn2}\\
 \ &=& \langle|\pi(\vec k_T)|^2\rangle/\omega(\vec k_T) \label{eq:gn3}.
\end{eqnarray}
Here $\omega(\vec k_T)$ is the eigenfrequency for a free massless plane wave
of the wave number $\vec k_T$ on a square lattice.
In our work, as in Lappi's, these prescriptions are applied to field 
configurations in the Coulomb gauge.
In Fig.~\ref{fig:dndkt} our transverse momentum distributions
for gluons are compared with Lappi's result.
KNV I (circles) in Fig.~\ref{fig:dndkt} is the same distribution
as in our previous papers, but with the correct value of $\Lambda_s$ used to
set the horizontal scale.
KNV I corresponds to the gluon number distribution from the definition
Eq.~(\ref{eq:gn1}),and $k_T$ dependence is extracted only for the transverse
momenta along the principal lattice directions.
The solid line is from Lappi's numerical result, which is computed
according to Eq.~(\ref{eq:gn3}) using the entire Brillouin zone.
KNV II is obtained by the same definition as Lappi's
and it is consistent. The deviation at large $k_T$ is considerd to be a consequence of 
how we define $k_T$ on the lattice. 
It should be noted that, with the lattice spacings used, the ultraviolet
portion of the spectrum, $k_T/\Lambda_s \geq 3$, cannot reliably reproduce
continuum physics.

The initial transverse energy per particle changes (because $f_E$ does)
 and is now 
\be
{E_T\over N} = 0.9 \Lambda_s \, .
\label{G3}
\ee
This value is nearly a factor of 2 lower than previously~\footnote{It is not exactly a factor of 
2 because the value of $\Lambda_s R$ at which $f_E$ is evaluated changes as well.}. As we will discuss 
in the following section, this lower value of the transverse energy per particle makes the connection 
to RHIC phenomenology simpler to interpret.

The ``formation time'' $\tau_i$ (defined as $\tau_i=1/\gamma(\Lambda_s R)/\Lambda_s$) of gluons
extracted in Ref.~\cite{AR00} (it is the same for SU(2) and SU(3) as confirmed in Ref.~\cite{AYR01}) is
unaffected, albeit the $\Lambda_s R$ at which it is evaluated should
be a factor of 2 larger than stated.  One has approximately
$\gamma=0.3$ in the range of interest. The initial energy density is then
\be
\varepsilon = {1\over \tau_i}\,{dE_T\over {\pi R^2 d\eta}} = {0.08\over g^2}\,\Lambda_s^4 \, .
\label{G4}
\ee

The number distributions,
 with coefficients appropriately scaled with $\Lambda_s$, 
have the same form as previously:
\begin{equation}
 {1\over \pi R^2}{dN\over d\eta d^2k_{T}}
     = {1\over g^2}{\overline f}_n(k_{T}/\Lambda_s)\, ,
\label{momdis}
\end{equation}
where ${\overline f}_n(k_{T}/\Lambda_s)$ is
\begin{equation}
{\overline f}_n = \left\{
  \begin{array}{ll}
     a_1\left[\exp\left(\sqrt{k_{T}^2+ m^2}/ T_{\rm eff}\right) -1\right]^{-1}
                       & (k_{T}/\Lambda_s \leq 1.5) \\ \\                

    a_2\,\Lambda_s^4\log(4\pi k_{T}/\Lambda_s)k_{T}^{-4}
                       & (k_{T}/\Lambda_s > 1.5) \\
\end{array} \right.
\label{eq:fit}
\end{equation}
with $a_1=0.137$, $m=0.0358\Lambda_s$, $T_{\rm eff}=0.465\Lambda_s$,
and $a_2=0.0087$.

\begin{figure}[t]
\includegraphics[width=3.3in]{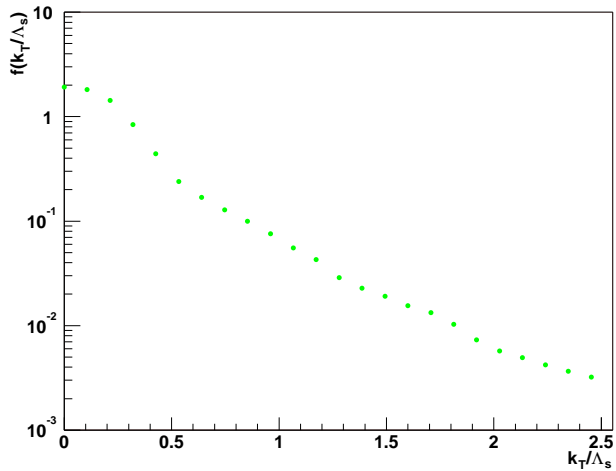}
\caption{Gluon occupation number defined in Eq~.(\ref{glue}), 
computed using Eq.~(\ref{eq:gn1}).
}
\label{fig:glue}
\end{figure}

With this expression for the number distributions, we can now compute
the occupation number for gluons after the collision.  The relevant
quantity for the validity of the three dimensional classical
approximation is the three dimensional occupation number~\footnote{We
thank Al Mueller for a discussion on this point.}  which can be
estimated from the two dimensional boost-invariant number distribution
computed in our simulation by the following
relation~\cite{MuellerQM2002}:
\be
f^{\rm glue} = {(2\pi)^3 \over 2\,(N_c^2-1)}\, {1\over \pi R^2}\,{dN\over d\eta d^2\,k_{T}}\,.
\label{glue}
\ee
Substituting our result for the number distribution from Eq.~(\ref{eq:gn1}) 
here,
one can compute the occupation number of gluons at the early stages of 
the collision. The results are shown in Fig.~\ref{fig:glue} for 
$\Lambda_s R=66$.
Strictly speaking, the classical
description is valid for $f > 1$.  It is not known however what the
value of $f$ is at which the classical description breaks down
completely and quantum corrections are dominant. 

All of our other results can be understood by appropriate
re-scaling. For instance, the $p_t$ distribution of the azimuthal
asymmetry, $v_2(p_t)$, plotted versus $p_t/\Lambda_s$ is the same
after replacing $\Lambda_s \rightarrow \Lambda_s/2$ on the x-axis of
the plot~\cite{AYR02a}.  The peak in $v_2(p_t)$ is therefore at even
lower momenta then previously -- at $p_t\sim \Lambda_s/8$.

\section{Numerical Gluodynamics and RHIC Phenomenology}

Now that we have shown that the results of two independent computations converge, it is useful to consider 
its implications for RHIC phenomenology. The results obtained in this classical approach are valid only at the 
very early stages of the nuclear collision and the connection to the final state distributions of measured 
hadrons might be considered remote. We will argue that, on the contrary, very general arguments can be 
applied to the RHIC data which will help constrain the properties of the initial state. In addition, they also 
significantly limit the freedom of final state interaction models which attempt to describe thermalization 
and the subsequent evolution of a Quark Gluon Plasma (QGP) in very high energy heavy ion collisions. 

We know that the hadronic multiplicity at $\eta \approx 0$ for central
Au-Au collisions with center of mass energy $\sqrt{s}=130$ GeV/nucleon
at RHIC is $\sim 1000$ while the hadronic transverse energy is $\sim
500$ GeV~\cite{experiment}. One expects, on general grounds, the following
relations, respectively, to hold for the transverse energy and the multiplicity:
\be
E_T^{\rm glue} &>& E_T^{\rm hadrons} \\
N^{\rm glue} &\leq& N^{\rm hadrons} \, .
\label{Constraint}
\ee
The inequality for the transverse energy follows from the expectation that some of the initial 
transverse energy is converted into work due to the expansion of the system~\cite{MuellerQM2002}. Moreover, if the 
system were to approach thermalization due to re-scattering, this process would again cause the 
initial transverse energy to be transferred into longitudinal energy~\cite{Mueller2000}. Even if 
there were no $pdV$ work done and no re-scattering, it is difficult to envision a scenario where the 
final transverse energy is larger than the initial one. The second constraint is also very plausible. 
For a thermal system, entropy is simply proportional to the multiplicity and 
the final entropy must be equal to or greater than the initial one. Even in the other extreme scenario, 
independent fragmentation with no re-scattering, parton-hadron duality suggests that the parton multiplicity prior to 
hadronization is equal to the multiplicity of hadrons after~\cite{phd}. 

Combining the first condition in Eq.~(\ref{Constraint}) on the transverse energy with Eq.~(\ref{G1}) and the 
empirical result for the RHIC data, we find that 
\be
\Lambda_s^{\rm RHIC}  > 1.3 \,\,{\rm GeV}\,.
\label{R1}
\ee
Here we have assumed that the average transverse area for the range of impact parameters corresponding to 
the most central collisions is $\pi R^2 = 130$ fm$^2$ and $g=2$~\footnote{There is clearly an ambiguity in this 
bound due to our choice of the strong coupling constant $g$, which cannot be resolved at this order of the 
computation. If we assume that $\alpha_s\equiv \alpha_s(\Lambda_s)$, the one loop value of $\alpha_s$ (from the 
bound we will set) is close to this estimate.}
Combining the second condition in Eq.~(\ref{Constraint}) on the particle multiplicity (at $\eta=0$) with 
Eq.~(\ref{G2}), we find 
\be
\Lambda_s^{\rm RHIC} \leq 2\,\, {\rm GeV}\,.
\label{R2}
\ee
Also, from Eq.~(\ref{G3}), and Eqs.~(\ref{R1})-(\ref{R2}),
 we obtain the following constraint for 
the initial transverse energy per particle:
\be
1.14 \,\,{\rm GeV} \leq {E_T\over N}_{\rm initial}^{\rm RHIC} \leq 1.76 \,\,{\rm GeV}\, ,
\label{R3}
\ee
and, for the initial energy density, (from Eq.~(\ref{G4})),
\be
7.1\,\, {\rm {GeV\over fm^3}} \leq \varepsilon_{\rm initial}^{\rm RHIC} \leq 40 \,\,{\rm {GeV\over fm^3}} \, .
\label{R4}
\ee
Thus the experimental data, the non-perturbative formulae
 in Eqs.~(\ref{G1})-(\ref{G2}) obtained from the 
classical numerical gluodunamics, and very general constraints can all be put together to significantly 
restrict the allowable range of $\Lambda_s^{\rm RHIC}$. These in turn restrict the initial transverse 
energy per particle and the initial energy density. How these quantities evolve tell us a great deal about 
the space-time evolution of partonic matter in a heavy ion collision. 

In our analysis, $\Lambda_s$ is an external parameter -- $\Lambda_s^2\equiv g^4\mu^2$, where $\mu^2$, as discussed 
previously, has the physical 
interpretation of being the average color charge squared per unit area per unit rapidity of color sources at 
higher values of $x$ than those of interest. This scale is not quite the same as the gluon saturation scale $Q_s^2$, 
which denotes the scale at which the gluon distribution changes qualitatively due to saturation effects~\cite{KovMuell}.
The relation between the two can be obtained by comparing the analytical expressions for the gluon distributions in 
the two approaches~\cite{JKMW,Kovchegov,KovMuell}. One 
obtains~\cite{AYR02b} 
\be
Q_s^2\approx \Lambda_s^2 \, {N_c\over 4\pi}\, \ln\left({\Lambda_s^2\over \Lambda_{QCD}^2}\right) \, .
\label{T1}
\ee
For the values of $\Lambda_s$ in the range of interest, $Q_s\approx \Lambda_s$. Extrapolating the Golec-Biernat--Wusthoff 
parametrization to RHIC (with the appropriate $A^{1/3}$ dependence), we find $\Lambda_s^2\approx Q_s^2 = 2$ GeV$^2$. 
This value for $Q_s$ is also obtained by Kharzeev and Nardi~\cite{KN}.
$\Lambda_s$ thus obtained lies within the range required by Eqs.~(\ref{R1})
 and Eq.~(\ref{R2}). Nevertheless, there is 
an ambiguity in Eq.~(\ref{T1}). For smaller values of $x$, one 
expects the infrared scale in this equation to be of order $Q_s$ and not $\Lambda_{QCD}$~\cite{KIIM,Mueller}; the relation 
between $Q_s$ and $\Lambda_s$ will then be modified. Secondly, even at the classical level, the requirement that 
color neutrality be maintained~\cite{Lam} will alter the relation between $\Lambda_s$ and $Q_s$. Indeed, in this case, one 
can directly relate the two scales with an appropriate definition of $Q_s$. This point will be discussed further at 
a later date. Understanding this relation is important because the constraint on $\Lambda_s$ discussed here will 
translate into a constraint on the gluon content in the nuclei at RHIC energies. 

The bounds in Eq.~(\ref{R3}) and Eq.~(\ref{R4}) are very important for
understanding the ``late'' stage dynamics of high energy heavy ion
collisions, namely, when the classical approach breaks down as it
must. One possibility is that the system undergoes hydrodynamic
expansion~\cite{Rasanen}. At the end of the classical
stage~\footnote{It is estimated in Ref.~\cite{BMSS} that the average
occupation number $f$, for $\alpha_s \ll 1$, is less than unity at $\tau
> 1/Q_s/{\alpha_s}^{3/2}$.}, the system is completely out of
equilibrium.  This is because the momentum distribution is very
anisotropic: $p_t\sim \Lambda_s$ and $p_z\sim 0$. If the system is to
thermalize, there must be particle production after the classical
stage. Particle number conserving $2\rightarrow 2$ interactions are
too inefficient to thermalize the system within a reasonable time
scale~\cite{Mueller2000,BMSS}. If $\Lambda_s$ is $2$ GeV, particles
cannot be produced subsequent to the classical stage because this
value saturates our bound for the particle number. If $\Lambda_s$ is
$1.3$ GeV (at the lower end of our allowed range), one can certainly
increase the particle number by approximately a factor of 2 after the
classical stage. The transverse energy per particle (see Eq.~(\ref{R3}))
will then also decrease by a commensurate amount and approach the
experimental value. However, one cannot then have any hydrodynamic
expansion (and therefore thermalization) because for $\Lambda_s=1.3$,
the transverse energy was already as low as it could be: hydrodynamic
expansion would lower it still further. 

The problem with this ``free
streaming'' scenario is that one does not have a satisfactory
mechanism to explain the momentum dependence of the azimuthal
anisotropy (especially at low $p_t$) observed at RHIC. The classical
numerical simulations of $v_2$ do not generate enough
$v_2$~\cite{AYR02a}. There is an interesting attempt to explain the
RHIC data as resulting from ``non-flow'' correlations but at low $p_t$
($p_t<Q_s$) its energy dependence is different from the trend in the
measured data from RHIC at $\sqrt{s}=130$ GeV/nucleon and
$\sqrt{s}=200$ GeV/nucleon~\cite{KovTuchin}. A consistent picture of
all the global features of the RHIC data might still be feasible
within the straitjacket provided by the classical simulations and the
data but there is limited room for models of final state interactions
to maneuver.

\begin{acknowledgments}
We would like to thank Rolf Baier, Keijo Kajantie, Dima Kharzeev, Tuomas
Lappi, Larry McLerran and Al Mueller for very useful discussions and
Tuomas Lappi and Keijo Kajantie for correspondence on their work.
R.~V.'s research was supported by DOE Contract No. DE-AC02-98CH10886
and the RIKEN-BNL Research Center.
Y.~N.'s research is supported by the DOE under Contracts DE-FG03-93ER40792.
 A.~K. and R.~V. acknowledge support
from the Portuguese FCT, under grants
POCTI/FNU/49529/2002 
and CERN/FIS/43717/2001 and support in part of NSF Grant No. PHY99-07949.
A.~K. and R.~V. also wish to thank the INT at the University of Washington
for partial support during the completion of this work.

\end{acknowledgments}

\end{document}